\documentclass[3p,times,procedia]{elsarticle}
\usepackage{nupha_ecrc}

\volume{00}

\firstpage{1}

\journalname{Nuclear Physics A}

\runauth{X. Guo, D. E. Kharzeev et al.}

\jid{nupha}

\jnltitlelogo{Nuclear Physics A}

\usepackage{amssymb}

\usepackage[figuresright]{rotating}

\newcommand{\md}{\mathrm{d}}

\begin{document}

\begin{frontmatter}



\dochead{XXVIth International Conference on Ultrarelativistic Nucleus-Nucleus Collisions\\ (Quark Matter 2017)}

\title{Chiral Vortical and Magnetic Effects in Anomalous Hydrodynamics}


\author[tsinghua]{Xingyu Guo\corref{cor1}}
\ead{xy-guo12@mails.tsinghua.edu.cn}
\author[SB,BNL]{Dmitri E. Kharzeev}
\author[fudan,fudan2]{Xu-Guang Huang}
\author[HUST]{Wei-Tian Deng}
\author[BNL]{Yuji Hirono}

\address[tsinghua]{Department of Physics, Tsinghua University and collaboration Innovation Center, Beijing, 100084, China}
\address[SB]{Department of Physics and Astronomy, Stony Brook University, Stony Brook, New York 11794-3800, USA}
\address[BNL]{Department of Physics, Brookheaven National Laboratory, Upton, New York 11973-5000, USA}
\address[fudan]{Physics Department and Center for Particle Physics and Field Theory, Fudan University, Shanghai 200433, China}
\address[fudan2]{The Key lab of Applied Ion Beam Physics, Ministry of Education, Fudan University, Shanghai 200433, China}
\address[HUST]{School of Physics, Huazhong University of Science and Technology, Wuhan 430074, China}
\cortext[cor1]{Corresponding author}

\begin{abstract}
We employ a 3+1D anomalous hydrodynamics with initial condition generated by HIJING to simulate the chiral vortical effect and the chiral magnetic effect in heavy-ion collisions. This allows us to calculate the charge-dependent two-particle correlations with respect to the reaction plane at different collision energies and centralities. We then compare the computed results with the experimental data and give discussions on the possible background effects.
\end{abstract}


\end{frontmatter}


\section{Introduction}
Chiral symmetry and its spontaneous breaking is one of the most important features of quantum chromodynamics. While in vacuum the chiral symmetry is broken, heavy ion collisions are able to create matter with high enough temperature and density so that the chiral symmetry is restored. In this way, they provide a precious chance to study unique phenomena related to chirality that are difficult to observe otherwise.

One of such phenomena is the anomalous chiral transport. Non-zero chiral chemical potential $\mu_5$, combined with an axial vector field such as magnetic field or vorticity, will produce a non-dissipative current\cite{KharzeevPLB633, KharzeevNPA797,  KharzeevNPA,FukushimaPRD78,DTSonPRD70,MetlitskiPRD72,CVEDubna}:
\begin{eqnarray}
\vec{J}_M&\propto&\mu_5 \vec{B},\\
\vec{J}_V&\propto& \mu_5 \mu \vec{\omega}.
\end{eqnarray}
These two effects are named chiral magnetic effect (CME) and chiral vortical effect (CVE), respectively. As these two currents can transport charge or baryon number, their occurrence in heavy-ion collisions will lead to charge or baryon number separation with respect to the reaction plane. In a single heavy ion collision event, the azimuthal distribution of observed particles can be decomposed into different Fourier components,
\begin{eqnarray}
\frac{\md N_\alpha}{\md \phi}&\propto&1+2v_{\alpha,1}\cos(\phi-\Psi_{RP})+2 v_{\alpha,2}\cos[2(\phi-\Psi_{RP})]+2a_{\alpha,1}\sin(\phi-\Psi_{RP})+\cdots,
\end{eqnarray}
with $v_{\alpha,n}$ the $n$th harmonic flow coefficient of particle spice $\alpha$. The chiral effects are supposed to contribute to $a_1$. However, as non-zero $\mu_5$ comes from fluctuation of axial charge, which changes sign in different events, the event average of $a_1$ will be zero. Thus we are forced to look for signals corresponding to $\langle a_1^2\rangle$. One of the most promising signals is the correlation\cite{VoloshinPRC78}:
\begin{eqnarray}
\gamma_{\alpha\beta}&=&\langle\cos(\phi^\alpha_1+\phi^\beta_2-2\Psi_{RP})\rangle,
\end{eqnarray}
with $\phi^{\alpha,\beta}_{1,2}$ is the azimuthal angle of the observed hadron. By taking the difference between the correlations of particles of the same charge (same-sign, SS) and opposite charge (opposite-sign, OS), $\langle a_1^2\rangle$ can be extracted, $\gamma_{OS}-\gamma_{SS}\sim \langle a_1^2\rangle$.

There have been ongoing searches for CME and CVE signals in heavy-ion collisions, with encouraging results\cite{STARPRL103,STARPRC81,STARPRC88,STARPRC89,STARPRL113,STARPRL110} measured. However, as there are complicated background effects, there are still debates on whether CME or CVE occurs in heavy-ion collisions (see the reviews\cite{KharzeevReview,LiaoReview,ROPP}. Therefore, it is necessary to develop a numerical framework where CME and CVE contributions can be calculated quantitatively which can be used for comparison with the experimental results. We would like to present a 3+1D relativistic anomalous hydrodynamics with both CME and CVE encoded which is developed on the bases of previous chiral magnetic hydrodynamics.\cite{HironoCMHD}.

\section{Hydrodynamics and Initial Condition}
The equations of motion for the anomalous hydrodynamics are:
\begin{eqnarray}
\partial_\mu T^{\mu\nu}&=&eF^{\nu}_{~\lambda}j_{e}^{\lambda},\\
\partial_\mu j_e^\mu&=&0,\\
\partial_\mu j_5^\mu&=&-C E^\mu B_\mu,
\end{eqnarray}
where $C$ is the anomaly constants, $E^\mu = F^{\mu\nu}u_\nu$, $B^\mu=\frac{1}{2}\epsilon^{\mu\nu\alpha\beta}F_{\alpha\beta}u^{\nu}$. We assume the fluid to be non-dissipative, so the energy-momentum tensor and currents are:
\begin{eqnarray}
T^{\mu\nu}&=&(\epsilon + p)u^\mu u^\nu -p\eta^{\mu\nu},\\
j_e^\mu&=&nu^\mu+\kappa_B B^\mu+\kappa_\omega  \omega^\mu,\\
j^\mu_5&=&n_{5}u^\mu+\xi_B B^\mu+\xi_\omega  \omega^\mu,
\end{eqnarray}
where
\begin{eqnarray}
e\kappa_B&\equiv& C\mu_{5}(1-\frac{\mu n}{\epsilon+p}),\\
e\xi_B&\equiv& C\mu(1-\frac{\mu_5 n_5}{\epsilon+p}),\\
e^2\kappa_\omega&\equiv& 2C\mu\mu_{5}(1-\frac{\mu n}{\epsilon+p}),\\
e^2\xi_\omega&\equiv& C\mu^2(1-\frac{2\mu_5 n_5}{\epsilon+p}),
\end{eqnarray}
are transport coefficients, with $\epsilon$ and $p$ are energy density and pressure, and $\eta^{\mu\nu} \equiv diag{(1,-1,-1,-1)}$. The second and third terms in each current correspond to the chiral effects. The transport coefficients are determined by requiring non-decreasing entropy\cite{SonTranCoef,KharzeevTranCoef}, or by using fluid-gravity duality\cite{KalaydzhyanTranCoef}.

We choose the equation of state to be the one for massless quark-gluon plasma with $\epsilon=3p$:
\begin{eqnarray}
p(T,\mu,\mu_{5})=\frac{g_{QGP}\pi^2}{90}T^4+\frac{N_cN_f}{6}(\mu^2+\mu_{5}^2)T^2 +\frac{N_cN_f}{12\pi^2}(\mu^4+6\mu^2\mu_{5}^2+\mu_{5}^4),
\end{eqnarray}
where $g_{QGP}=g_g+\frac{7}{8}g_q$, $g_g=(N^2_c-1)N_s$, $g_q=2N_cN_sN_f$; $N_c=3$, $N_f=3$ and $N_s=2$ are the number of colors, flavors and spin states.

On the freezeout hypersurface with $T=160$ MeV, Monte-Carlo sampling of hadrons based on standard Cooper-Frye approach is performed.

The magnetic field is taken as a background field in the direction proportional to the reaction plane\cite{KharzeevNPA} and takes the form\cite{HironoCMHD}:
\begin{eqnarray}
eB_y(\tau,\eta_s,x,y)=eB_0\frac{b}{2R}exp\left(-\frac{x^2}{\sigma^2_x}-\frac{y^2}{\sigma^2_y}-\frac{\eta_s^2}{\sigma^2_\eta}-\frac{\tau}{\tau_B}\right).
\end{eqnarray}

In order to get correct fluid evolution, initial condition is crucial. As the vorticity in the system comes mainly from the initial collision, a model that provides non-trivial velocity distribution is necessary. The HIJING model has been used to describe the vorticity of initial system successfully\cite{DengHIJING}; on the other hand, HIJING does nor provide axial charge distribution, which is also necessary to CME and CVE. As a compromise, we first calculate the event average of $\gamma$ correlation with only CME and extended MC-Glauber model\cite{AlverGlauber}. Then, we use HIJING model to introduce initial vorticity and assume $\mu_5/T = 0.1$ to be a constant and compare the contribution from CME and CVE.

\section{Numerical Results}
\begin{figure}[h]
	\begin{center}
	\includegraphics[width = 0.45\textwidth]{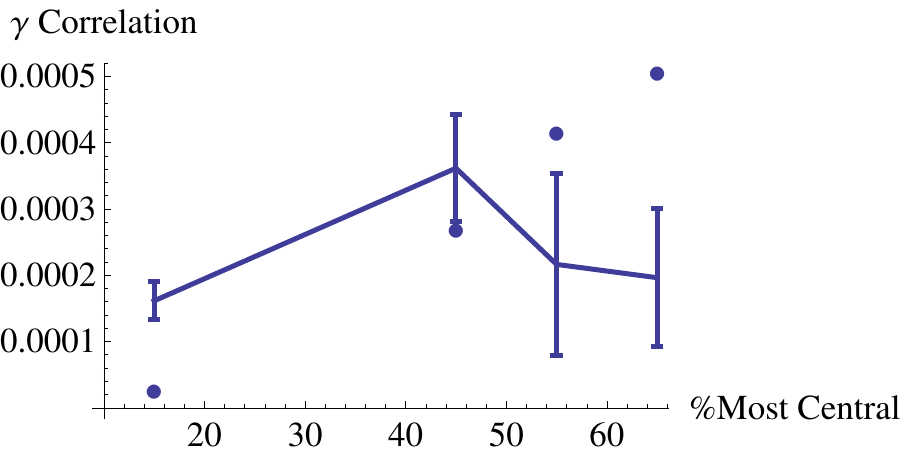}
	\includegraphics[width = 0.45\textwidth]{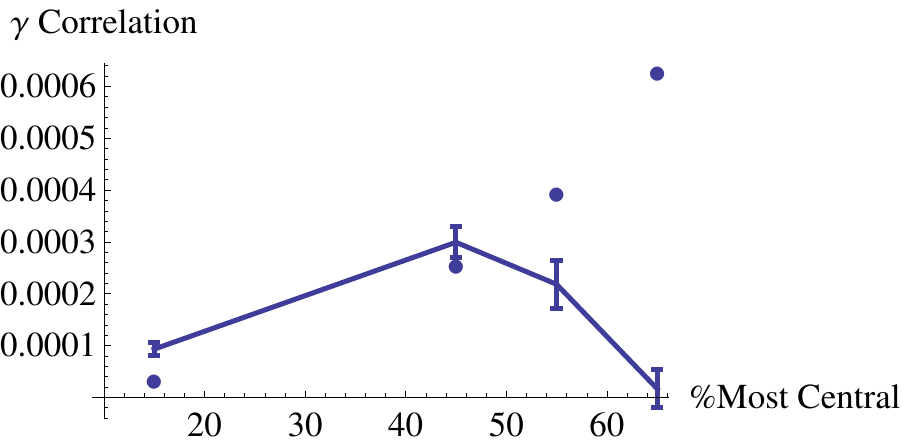}
	\caption{p-$\pi$ (left) and $\pi-\pi$ (right) OS-SS correlation (line) compared with experimental data (dots)}
	\label{correlation}
	\end{center}
\end{figure}

Until now, most of theory calculation and experimental search have been focused on the correlation of all charged particles. More information could be extracted by looking into the correlation of different particle species. Also there have been results of correlation of identified particles at RHIC 200 GeV Au-Au collisions\cite{WenData}. Thus we calculated the corresponding correlation with CME contribution and made a comparison. The results are shown in Fig.\ref{correlation}. The calculation results match data roughly with no calibration of parameters at low centralities. We clearly underestimated the correlation for more peripheral collisions, where CVE are supposed to become more important.

\begin{table}[h]
	\caption{Pion and proton $a_1$ with different effects included, for RHIC 200Gev Au-Au $40\sim50\%$ centrality collisions}
	\begin{center}
	\begin{tabular}{|c|c|c|}
	\hline
	Contribution & $\langle a_1^\pi\rangle$ & $\langle a_1^p\rangle$\\
	\hline
	None & $-0.0023\pm0.0014$ & $-0.0120\pm0.0038$\\
	CME & $0.0013\pm0.0013$ & $-0.0027\pm0.0042$\\
	CME + CVE & $0.0048\pm0.0018$ & $0.0109\pm0.0044$\\
	None (Glauber initial condition) & $-0.0007\pm0.0019$ & $-0.0058\pm0.0063$\\
	\hline
	\end{tabular}
	\end{center}
	\label{a1}
\end{table}

As described above, we made a comparison between CME and CVE contributions at $40\sim 50\%$ centrality, 200 Gev Au-Au collisions. As the axial charge are taken as an arbitrary constant, it may not be suitable to compare with experimental data, however in this case $a_1$ will not change sign at different events so we can calculate its average directly. Table \ref{a1} shows results with different contributions. We calculated the averaged $a_1$ with no chiral effects, only CME and both CME and CVE. We also calculated $a_1$ with Glauber initial condition and no chiral effect as control. The background fluctuation from HIJING model is larger than Glauber model. Taking this into consideration, we can see that even at middle centrality, the contribution from CVE and CME are comparable. We also notice that the average $a_1$ for protons is larger than pions, which is also consistent with experimental results.

\section{Conclusion}
Using a 3+1D relativistic anomalous hydrodynamic model, we calculated the $\gamma$ correlation of identified particles for RHIC 200 Gev Au-Au collisions. The results are comparable to experimental in middle centrality regions. We also made a comparison between CME and CVE and found that these two effects are of the same order of magnitude. More results will be reported in future.

Acknowledgements: The author would like to thank Shuzhe Shi and Yi Yin for very helpful discussions. X. G is supported by Chinese Scholarship Council.




\bibliographystyle{elsarticle-num}
\bibliography{CMEbib}

\begin{thebibliography}{10}
\expandafter\ifx\csname url\endcsname\relax
  \def\url#1{\texttt{#1}}\fi
\expandafter\ifx\csname urlprefix\endcsname\relax\def\urlprefix{URL }\fi
\expandafter\ifx\csname href\endcsname\relax
  \def\href#1#2{#2} \def\path#1{#1}\fi

\bibitem{KharzeevPLB633}
D.~E. Kharzeev, Phys. Lett. B 633 (2006) 260.

\bibitem{KharzeevNPA797}
D.~E. Kharzeev, A.~Zhitnitsky, Nucl. Phys. A 797 (2007) 67.

\bibitem{KharzeevNPA}
D.~E. Kharzeev, L.~D. McLerran, H.~J. Warringa, Nucl. Phys. A 803 (2008) 227.

\bibitem{FukushimaPRD78}
K.~Fukushima, D.~E. Kharzeev, H.~J. Warringa, Phys. Rev. D 78 (2008) 074033.

\bibitem{DTSonPRD70}
D.~T. Son, A.~R. Zhitnitsky, Phys. Rev. D 70 (2004) 074018.

\bibitem{MetlitskiPRD72}
M.~A. Metlitski, A.~R. Zhitnitsky, Phys. Rev. D 72 (2005) 045011.

\bibitem{CVEDubna}
O.~Rogachevsky, A.~Sorin, O.~Teryaev, Phys. Rev. C 82 (2010) 054910.

\bibitem{VoloshinPRC78}
S.~A. Voloshin, Phys. Rev. C 70 (2004) 057901.

\bibitem{STARPRL103}
B.~I. Abelev, et~al., Phys. Rev. Lett. 103 (2009) 251601.

\bibitem{STARPRC81}
B.~I. Abelev, et~al., Phys. Rev. C 81 (2010) 054908.

\bibitem{STARPRC88}
L.~Adamczyk, et~al., Phys. Rev. C 88 (2013) 064911.

\bibitem{STARPRC89}
L.~Adamczyk, et~al., Phys. Rev. C 89 (2014) 044908.

\bibitem{STARPRL113}
L.~Adamczyk, et~al., Phys. Rev. Lett. 113 (2014) 052302.

\bibitem{STARPRL110}
L.~Adamczyk, et~al., Phys. Rev. Lett. 110 (2013) 021301.

\bibitem{KharzeevReview}
D.~E. Kharzeev, J.~Liao, S.~A. Voloshin, G.~Wang, Prog. Part. Nucl. Phys. 88
  (2016) 1.

\bibitem{LiaoReview}
J.~Liao, Pramana 84 (2015) 901.

\bibitem{ROPP}
X.-G. Huang, Rep. Prog. Phys. 79 (2016) 076302.

\bibitem{HironoCMHD}
Y.~Hirono, T.~Hirano, D.~E. Kharzeev, arXiv:  1412.0311.

\bibitem{SonTranCoef}
D.~T. Son, P.~Surowka, Phys. Rev. Lett. 103 (2009) 191601.

\bibitem{KharzeevTranCoef}
D.~E. Kharzeev, H.-U. Yee, Phys. Rev. D 84 (2011) 045025.

\bibitem{KalaydzhyanTranCoef}
T.~Kalaydzhyan, I.~Kirsch, Phys. Rev. Lett. 106 (2011) 211601.

\bibitem{DengHIJING}
W.-T. Deng, X.-G. Huang, Phys. Rev. C 93 (2016) 064907.

\bibitem{AlverGlauber}
B.~Alver, M.~Baker, C.~Liozides, P.~Steinberg, arXiv:  0805.4411.

\bibitem{WenData}
L.~Wen, in: The XXVI international conference on ultrarelativistic heavy-ion
  collisions (Quark Matter 2017), 2017.

\end{thebibliography}







\end{document}